\documentclass{IOS-Book-Article}
\usepackage[hide]{ed}
\usepackage{local}

\begin{document}
\pagestyle{headings}
\def\thepage{}

\makeatletter
\setlength\columnsep{4mm}
\setlength\intextsep   {10\p@ \@plus 6\p@ \@minus 4\p@}
\makeatother

\begin{frontmatter}              
\title{Context Graphs for Legal Reasoning and Argumentation}

\markboth{}{January 2020\hb}

\author[A]{\fnms{Max} \snm{Rapp}},
\author[A]{\fnms{Axel} \snm{Adrian}}
and
\author[A]{\fnms{Michael} \snm{Kohlhase}}

\address[A]{FAU Erlangen-N\"urnberg}

\begin{abstract}
  We propose a new, structured, logic-based framework for legal reasoning and argumentation:
  Instead of using a single, unstructured meaning space, theory graphs organize knowledge and inference into collections of modular meaning spaces organized by inheritance and interpretation. 
  Context graphs extend theory graphs by attack relations and interpret theories as knowledge contexts of agents in  argumentation.
  We introduce the context graph paradigm by modeling the well-studied case Popov v. Hayashi, concentrating on the role of analogical reasoning in context graphs. 
\end{abstract}

\end{frontmatter}
\markboth{January 2020\hb}{January 2020\hb}

\section{Introduction}\label{sec:intro}
There is a huge literature in legal theory regarding the methods of jurisprudence; see e.g. \cite{Ashley:aila17,Blumenwitz:eaar03,Hart:tcl12,MacCormick:lrlt78}. But there is no consensus on how legal thinking actually works or should work in detail. One area of dissent is between rule-based and analogical reasoning when discussing precedent in case law systems \cite{Stevens2018:RulesvAnalogies}.
However, it seems all approaches structurally combine the legal rule of a case -- especially of a leading case -- and the legally important facts of the present case \cite[p. 879]{Adrian:gjfm09}.

This structural combination of laws and facts can be represented in formal systems as a logical syllogism: for each case to be decided it has to be checked in detail whether the legally relevant facts can be ``subsumed'' under the conditions of an applicable legal rule. See e.g. \cite[p. 19ff, 229ff]{MacCormick:lrlt78}, \cite[p. 124ff]{Hart:tcl12}, or \cite[p. 777ff]{Adrian:gjfm09} for a discussion. This is sometimes \cite[p.23ff]{Engisch:ejd97} seen as establishing the analogical equivalence of the present case and a legally relevant precedent.

\paragraph{Contribution}
We propose to capture these analogical modes of reasoning in a new, structured, logic-based framework for legal reasoning and argumentation: context graphs. These structure legal and world knowledge into theories, which are organized into a strongly interconnected graph by inheritance, interpretation, and argumentation relations. This graph structure drives knowledge processing and argumentation.

\paragraph{Overview}
\Cref{sec:soa} reviews the relevant state of the art in legal reasoning. \Cref{sec:ccgraphs} introduces the context graphs framework and applies it to the \emph{Popov v. Hayashi} case. \Cref{sec:pushout} shows rule and precedence case application in the case and \cref{sec:analogy} discusses the fine points of analogical transfer in these applications. \Cref{sec:concl} concludes the paper. 

\paragraph{Acknowledgements} The work presented in this paper was supported by the DFG SPP 1999 RATIO (Rational Argumentation Machines) under Grant KO 2428/18.

\section{State of the Art in Modelling Legal Reasoning with Cases}\label{sec:soa}

Capturing and automating how law practitioners reason with cases has been a major focus in legal \AI since the \HYPO \cite{Ashley1990:HYPO} and \CATO \cite{Aleven1997:CATO} systems of the late 80s and early 90s. There are two main approaches in modelling legal reasoning with cases: rule-based and analogy-based reasoning \cite{Stevens2018:RulesvAnalogies,AtkinsonBenchCapon2019:RulesvAnalogies}.

In \emph{analogical} approach the similarity of cases is assessed based on \emph{aspects}. That is, \emph{dimensions} of cases that range from maximally pro-defendant to maximally pro-plaintiff and boolean \emph{factors} that pertain either to the defendant or to the plaintiff.
Conflicting arguments for different possible precedent applications are then constructed based on the degree to which cases resemble each other in terms of their aspects \cite{CCapon2012:PvHFactorsDimensions,AtkinsonBenchCapon2019:RulesvAnalogies}. The main implementations of the analogical approach are the \HYPO \cite{Ashley1990:HYPO} and \CATO systems \cite{Aleven1997:CATO}.
In the \emph{rule-based} approach, precedents are taken to extend the existing body of law by their respective ``rule of the case'' which is encoded as a defeasible inference rule in the object language of some logic (e.g. \cite{Prakken:PvHaspic+}). Defeasibility is established through attacks (undercuts, rebuttals) and preference ordering over rules.
The most important systems include \aspic \cite{PrakkenModgil2014:aspic+,Prakken:PvHaspic+} and abstract dialectical frameworks (\adfs) in conjunction with the \angelic methodology \cite{Al-Abdulkarim2016:ADFsCaseLaw}.

In recent years, rule-based approaches have proven to be easier to operationalize as they mostly require checking whether rules have been correctly applied. Another advantage of rule-based approaches is that they allow to systematically treat the \emph{burden of proof} and different \emph{proof standards} through the use of default reasoning and prioritisation of rules respectively.

However, \cite{Stevens2018:RulesvAnalogies} argues that the analogy-account for legal reasoning is preferable as it better accounts for judicial discretion and as a heuristic to find applicable precedents as well as reasons to distinguish a case -- especially for novel cases.
\cite{AtkinsonBenchCapon2019:RulesvAnalogies} concur with this analysis but conclude that computational models for analogical reasoning are unlikely: they would require huge ontologies marred by the problem of subjectivity and the impossibility to choose between competing ontologies of the same set of facts. They do however see potential for analogy-based tools for training law-students through ex-post analysis and assisting law-practitioners in constructing ontologies for novel cases piecemeal (see e.g. \cite{Ashley2009:legalOntologies}).
In this light, we believe that checking the internal coherence of analogical reasoning, automating some of its aspects and guiding practitioners through heuristic detection of analogy candidates is in the realm of the feasible. See \cref{sec:pushout,sec:analogy} for a discussion. 



\section{Formalizing the Law in Context Graphs}\label{sec:ccgraphs}
In the rule-based approach legal reasoning is represented as inference in logical calculi. To this end classical logical calculi are extended by defeasible inference rules or default assumptions in various rule-based or assumption-based argumentation frameworks -- e.g. \aspic \cite{PrakkenModgil2014:aspic+}or \aba \cite{CyrasFanSchulzToni2017:ABA}.

In this paper, we adopt this practice, but introduce an important novel feature: instead of using a single, unstructured meaning space, we organize the knowledge, inference and argumentation into collections of meaning spaces we call \textbf{theories} or \textbf{contexts} when taking argumentation into account.
Importantly, theories are not independent, but are connected by inheritance and interpretation morphisms, forming theory graphs. This enables us to combine rule-based and analogical reasoning through theory morphisms. For argumentation, these morphisms can be used to define conflict relations like attack.

\paragraph{Theory/Context Graphs in \mmt}\label{sec:mmt}

For concrete representation of the theory graphs we use the  \mmt (Meta Meta Theories) format \cite{RabKoh:WSMSML13}.It models formal objects and statements using logical frameworks, in particular the judgments-as-types paradigm.

Formally, theories are sets of object declarations and definitions as well as statements (axioms and theorems).
Theory graphs are diagrams in the categories of theories and morphisms.\ednote{MR@MK: category?}
The former allows for fine-grained specifications of the semantics of individual objects, and the latter allows for inducing and translating knowledge across theories.

The possible morphisms in \mmt are \textbf{inclusions}, which import all declarations from the domain to the co-domain, \textbf{structures}, which are like includes but copy and translate all declarations, \textbf{views}, which are semantics-preserving translations from domain to codomain, and the \textbf{meta-theory}-relation, which behaves like an include for most purposes.
We refer the reader to \cite{RabKoh:WSMSML13} for the general theory of theory graphs and~\cite{KohRapp:cgal19} for the idea of context graphs; we will introduce the features that are relevant to legal reasoning using the examples as we go along.

\begin{wrapfigure}l{8.3cm}\vspace*{-.5em}
  \sf
\begin{tikzpicture}[xscale=.95]
  \node[thy] (l) at (0,0) {\begin{tabular}{c}World\\ Knowledge\\+\\Legal\\ Foundation\end{tabular}};

  \node[thy] (kvcR) at (-3,-1) {KvC-Rule};
  \node[thy] (kvcF) at (-3,0) {KvC-Aspects};
  \node[thy] (kvcL) at (-3,1) {KvC-Lexicon};
  \draw[include] (kvcL) -- (kvcF);
  \draw[include] (kvcF) -- (kvcR);
  \draw[include] (l) -- (kvcL);

  \node[thy] (pvhF) at (3,-2) {PvH-Facts};
  \node[thy] (pvhL) at (0,-2) {PvH-Lexicon};
  \draw[include] (pvhL) -- (pvhF);
  \draw[include] (l) -- (pvhL);

  \node[thy] (mvsR) at (3,-1) {MvS-Rule};
  \node[thy] (mvsF) at (3,0) {MvS-Aspects};
  \node[thy] (mvsL) at (3,1) {MvS-Lexicon};
  \draw[include] (mvsL) -- (mvsF);
  \draw[include] (mvsF) -- (mvsR);
  \draw[includeleft] (l) -- (mvsL);

  \node (gr)  at (-3,3.6) {\rm\footnotesize Gray's Rule};
  \node[thy] (grC)  at (-3,2.2) {Gray-Cond};
  \node[thy] (grR)  at (-3,3.2) {Gray-Rule};
  \node[draw,fit=(gr) (grC),rounded corners] (grg) {};
  \draw[include] (grC) -- (grR);
  \draw[include] (l) -- (grC);

  \node (df)  at (0,3.6) {\rm\footnotesize Default Rule};
  \node[thy] (dfC)  at (0,2.2) {Default-Cond};
  \node[thy] (dfR)  at (0,3.2) {Default-Rule};
  \node[draw,fit=(df) (dfC),rounded corners] (dfg) {};
  \draw[include] (dfC) -- (dfR);
  \draw[include] (l) -- (dfC);

  \node (mc)  at (3,3.6) {\rm\footnotesize McCarthy's Rule};
  \node[thy] (mcC)  at (3,2.2) {McCarthy-Cond};
  \node[thy] (mcR)  at (3,3.2) {McCarthy-Rule};
  \node[draw,fit=(mc) (mcC),rounded corners] (mcg) {};
  \draw[include] (mcC) -- (mcR);
  \draw[includeleft] (l) -- (mcC);

  \draw[thick,red,rounded corners] ([xshift=-2pt,yshift=2pt]grg.north west)
      -- ([xshift=2pt,yshift=2pt]mcg.north east)
      -- ([xshift=2pt,yshift=-2pt]mcg.south east)
      -- ([xshift=5pt,yshift=10pt]l.north east)
      -- ([xshift=5pt,yshift=-8pt]l.south east)
      -- ([xshift=-5pt,yshift=-8pt]l.south west)
      -- ([xshift=-5pt,yshift=10pt]l.north west)
      -- ([xshift=-2pt,yshift=-2pt]grg.south west)
      -- cycle;

   \node[draw,thick,green,rounded corners,fit=(kvcL) (kvcR) (l)]{};
   \node[draw,thick,blue,rounded corners,fit=(mvsL) (mvsR) (l),inner sep=1pt]{};

   \draw[thick,cyan,rounded corners] ([xshift=-8pt,yshift=8pt]l.north west)
      -- ([xshift=8pt,yshift=8pt]l.north east)
      -- ([xshift=8pt,yshift=-8pt]l.south east)
      -- ([xshift=2pt,yshift=2pt]pvhF.north east)
      -- ([xshift=2pt,yshift=-2pt]pvhF.south east)
      -- ([xshift=-11pt,yshift=-2pt]pvhL.south west)
      -- cycle;

\end{tikzpicture}\vspace*{-.5em}
  \caption{\pvh as a Context Graph}\label{fig:SubsumtionGraph}\vspace*{-.5em}
\end{wrapfigure}
The \textbf{meta-theory} relation links a logical framework to the logics defined in it, thus formalizing the ``logics-as-theories'' approach, which allows us to represent logics and (legal) inference systems together with the domains in the \mmt system.
This is an integral part of the utility of the \mmt  system for our approach, but we will gloss over this aspect in this paper, concentrating on the domain-level structure of legal reasoning and argumentation. 

The \omdoc/\mmt language is implemented in the \mmt system (Meta Meta Toolset; see~\cite{uniformal:URL}), which provides an API for the language constructs at all levels and provides both logical services such as type reconstruction and rewriting and knowledge management services such as IDE and HTML presentation and browsing of libraries.

\paragraph{Running Example: Popov vs. Hayashi}\label{sec:running}

We will use the seminal case ``Popov vs. Hayashi''\cite{Atkinson2012:PopovvHayashiRuling} -- reduced to the salient argumentative backbone -- as a running example in this paper. The full \mmt formalization of the example is available at \cite{popovhayashi:url}. 

The black part in \Cref{fig:SubsumtionGraph} shows a theory graph of legal knowledge. In the center we have a background ``theory'' of general legal and world knowledge, which we presuppose here.
(In reality, this is  -- of course -- a large, modular theory graph, which is shared and selectively imported by particular formalizations.)
Clustered around this are formalizations of particular legal knowledge: various precedents and legal rules, which are theory graphs on their own (see \Cref{sec:cases-rules}) and inherit from the background theory via inclusion morphisms (\mmtar{include} in the diagrams in this paper).


 Let us now consider the advantages of a modular theory graph over a representation as an (unstructured) collection of formulae: The first division we can encode concerns the sources from which the text draws. In \cref{fig:SubsumtionGraph} the sources are represented in a context graph. Typically the distinction is threefold in case law applications:
\begin{itemize}
\item\textbf{Precedents}: A collection of already decided cases. Their opinions give \emph{concrete} descriptions of the \emph{aspects} that contributed to the \emph{ratio} of the decision. A rule established by a precedent x is called a \emph{rule of x}. \cref{fig:SubsumtionGraph} contains two precedents cited in \pvh grouped in green/blue: \kvc and \mvs.
\item \textbf{Rules}: Some rules are laid down in an \emph{abstract} form, for example in legal commentary. In \cref{fig:SubsumtionGraph} the rule subgraph (grouped in red) contains Gray's rule which Judge McCarthy adopted; a default rule (cf. \cref{sec:A3}); and the rule of \pvh fashioned by Judge McCarthy himself.
\item \textbf{Facts}: These are the facts of the case at hand. In \cref{fig:SubsumtionGraph} they are grouped in cyan. We assume the facts to be settled. The problem of evidence valuation could potentially be addressed with similar methods as we propose here - however we leave this to future work.
\end{itemize}

Note that all of these include the world/legal background, and thus the various groups overlap in that.
The shared background makes the formalizations coherent.
In formalization practice this means that all shared or shareable knowledge needs to move into the background. 

The subsumption task the court faces is to establish that the \pvh facts instantiate the aspects allowing for the application of the rule of \kvc.
To this end, the facts of \pvh have to be shown to be analogous to the legally relevant aspects of \kvc with the help of information from the other subgraphs.

\paragraph{Objects, Statements, and Proofs in \mmt}

\begin{wrapfigure}r{4cm}\vspace*{-1em}
  \includegraphics[width=4cm]{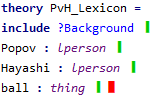}\vspace*{-.5em}
  \caption{An \mmt Theory}\label{fig:surface}\vspace*{-1.5em}
\end{wrapfigure}
Before we can take a look at the contents of the theories shown above, we have to discuss the representation of objects and statements.
In \mmt we can formalize then as \textbf{declarations} of the form
\begin{quote}\sf
  name : type = definiens
\end{quote}
where \cn{name} is the object name, (optionally) \cn{type} its type, and (again optionally) \cn{definiens} can be used to define the object.

\Cref{fig:surface} shows a snippet of our formalization in \mmt surface syntax -- we will us a slightly simplified version of that in the theory graph diagrams below.
The theory \cn{PvH-Lexicon} contains an inclusion of the theory \cn{background} -- the \cn{?} is the syntactic indicator for theory names -- and three object declarations: the key contributors to our example case.  The coloured rectangles are \mmt delimiters of various levels.

In the judgments-as-types paradigm \mmt employs, we can formalize statements in the same way.
The main trick is to introduce a type constructor $\vdash$, which -- given a proposition $\phi$ -- constructs the type $\vdash\phi$ of all proofs of $\phi$.
If $\vdash\phi$ is non-empty, then $\phi$ has a proof, and thus $\phi$  is a theorem.
Thus a declaration \cn{c : $\vdash\phi$} which declares a constant \cn{c} of type $\vdash\phi$ entails that $\vdash\phi$ is non-empty, and can be seen as an axiom.
Note that this trick also allows us to formalize inference rules in a calculus, e.g. the conjunction elimination $\wedge E$ rule of the natural deduction calculus as a function $\vdash\phi\wedge\psi\;\rightarrow\;\;\vdash\phi\;\rightarrow\;\;\vdash\psi$.
With such rules (and axioms like \cn{c} above) proofs become terms \mmt and theorems become declarations, e.g. $\cn{pf} : \vdash\phi = \wedge E\;\cn{c}\; \cn{c}$ which defines a constant \cn{pf} of type $\vdash\phi$ by the proof $\wedge E\;\cn{c}\;\cn{c}$ from the axiom \cn{c : $\vdash\phi$} discussed above (used twice).

We will use a meta-theory \cn{FOLND} formalizing first-order logic and its natural deduction calculus which introduces the constants $\rightarrow$, $\vdash$, $\wedge$, $\neg$, $\forall$, etc. in the rest of the paper.
All the theory/context graph diagrams should include a node for this theory with a meta link to all other theories, but we omit these for simplicity.  

In \cref{sec:A3} we will discuss context graphs for argumentation. There we graduate to graphs that include argumentation relations as special morphisms and a meta-theory for defeasible aspects of legal reasoning.

\paragraph{Formalizing Precedents, Rules and Facts}\label{sec:cases-rules}

Rules map sets of legal conditions to sets of legal consequences. One could capture this characteristic of rules at the object-language level as is done in rule based argumentation approaches. However to enable analogical reasoning we opt for a \emph{theory-graph-based} approach to rules. We regard the legal conditions and consequences as separate theories connected by a theory morphism.


Given a meta-theory, for a formalization of a rule we thus need 

\begin{compactenum}\label{norm:requirements}
\item a legal lexicon of constants specifying the concepts in the rule;
\item a theory of legal conditions specifying when the norm applies;
\item a theory of legal consequences specifying what follows from the rule application.
\end{compactenum}

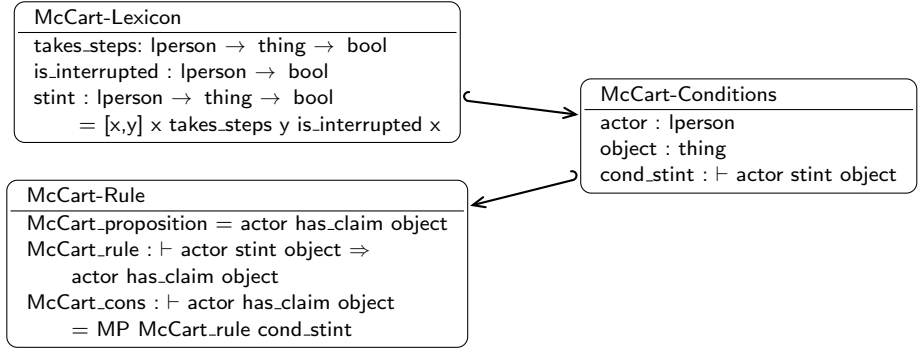
\begin{figure}
  \begin{tikzpicture}[xscale=1.7,yscale=1.7]\sf\footnotesize
  \node[thy] (l) at (0,0)
  {\begin{tabular}{l}
     McCart-Lexicon\\\hline
     takes\_steps: lperson $\ar$ thing $\ar$ bool\\
     is\_interrupted : lperson $\ar$ bool\\
     stint : lperson $\ar$ thing $\ar$ bool\\
     \qquad = [x,y] x takes\_steps y is\_interrupted x
   \end{tabular}};

  \node[thy] (c) at (4,-.5)
  {\begin{tabular}{l}
     McCart-Conditions\\\hline
     actor : lperson\\
     object :  thing\\
     cond\_stint : $\vdash$ actor stint object\\
   \end{tabular}};
 
  \node[thy] (r) at (0,-1.5)
  {\begin{tabular}{l}
     McCart-Rule\\\hline
     McCart\_proposition = actor has\_claim object\\
     McCart\_rule : $\vdash$ actor stint object $\Rightarrow$\\
     \qquad actor has\_claim object\\
     McCart\_cons : $\vdash$ actor has\_claim object \\
     \qquad = MP McCart\_rule cond\_stint
   \end{tabular}};
 \draw[include] (l) -- (c); 
 \draw[includeleft] (c) -- (r); 
\end{tikzpicture}
  \caption{A rule graph for the rule judge McCarthy introduces in his opinion.}\label{fig:normgraph}
\end{figure}

\Cref{fig:normgraph} shows a rule formalization in three theories for the three concerns listed above. 
General concepts such as the types ``lperson''  and ``thing'' are inherited from the background knowledge.
The lexicon contains the concepts that are to be defined in the rule. In this case, the predicates ``takes\_steps'', ``is\_interrupted'' and their conjunction ``stint''.
Condition theories declare a set of conditions necessary for the application of the rule.
E.g. McCarthy's rule as depicted in \cref{fig:normgraph} requires an actor and an object and has the condition-axiom \textsf{cond\_stint} requiring that the actor takes steps to take possession of the object and is interrupted.
Note that this is similar to approaches using abstract dialectical frameworks \cite{Al-Abdulkarim2016:ADFsCaseLaw}. We leave an exploration this connection to future work.

Finally, the theory ``McCart-Rule'' contains three declarations: a proposition to which the rule pertains; an object language version of the rule itself; and finally the legal consequences of the rule application. While the rule is declared axiomatically, the legal consequences come with a derivation from the condition axioms and the rule axiom.
In \cref{par:precedent} we will see how this enables us to ``inject'' an instantiation of the rule as well as a proof of the legal consequences into the present case where it can then be shared and translated across theories. 

Formalization of precedents is very similar to formalization of rules.
The main difference is that in precedent graphs conditions are replaced by \emph{aspects} -- not all of which may be necessary conditions for application of the rules that the precedent establishes.

Finally, in the present case, it is not apparent yet which facts of the case will be legally relevant and become aspects. Thus in a \emph{present case graph} the legal conditions theory is replaced with a theory describing the case's facts.




\section{Precedent/Rule Application via Views and Pushouts}\label{sec:pushout}
We have seen the basic ``object-oriented''  infrastructure of theory graphs above and how legal knowledge representation can profit from modularity.
But for legal reasoning we utilize an advanced feature of \mmt: \textbf{views} and \textbf{pushouts}.
In this section we use them to combine modularity and inference to model analogical reasoning.

\begin{figure}[ht]
  \sf
\begin{tikzpicture}[xscale=1.7]
  \node[thy] (kvcR) at (0,0) {KvC-Rule};
  \node[thy,thick,blue] (pvhR) at (1.8,0) {PvH-Ruling};
  \node[thy] (kvcA) at (0,1) {KvC-Aspects};
  \node[thy,thick,blue] (pvhAM) at (1.8,1.5) {PvH-Asp-McCart};
  \draw[view,thick,blue] (kvcR) -- (pvhR);
  \draw[view] (kvcA) -- (pvhAM);
  \draw[include] (kvcA) -- (kvcR);
  \draw[include,thick,blue] (pvhAM) -- (pvhR);
  \textcolor{blue}{\sepushout[.7]{kvcA}{pvhR}}

  \node[thy] (mcR) at (0,2) {McCart-Rule};
  \node[thy] (mcC) at (0,3) {McCart-Cond};
  \node[thy,thick,blue] (pvhAG) at (3,2.5) {PvH-Asp-Gray};
  \draw[view,thick,blue] (mcR) -- (pvhAM);
  \draw[view] (mcC) to[bend right=23] (pvhAG);
  \draw[include] (mcC) -- (mcR);
  \draw[include,thick,blue] (pvhAG) -- (pvhAM);
  \textcolor{blue}{\sepushout[.8]{mcC}{pvhAM}}

  \node[thy] (pvhA) at (3,3.5) {PvH-Facts};
  \node[thy] (grC) at (1.5,3.8) {Gray-Cond};
  \node[thy] (grR) at (1.5,2.8) {Gray-Rule};
  \draw[include,thick,blue] (pvhA) -- (pvhAG);
  \draw[include] (grC) -- (grR);
  \draw[view] (grC) -- (pvhA);
  \draw[view,thick,blue] (grR) -- (pvhAG); 
  \textcolor{blue}{\sepushout[.8]{grC}{pvhAG}}

  \node[thy] (dfR) at (6,2) {Default-Rule};
  \node[thy] (dfC) at (6,3) {Default-Cond};
  \node[thy,thick,blue] (pvhAD) at (4.2,1.5) {PvH-Asp-Default};
  \draw[view,thick,blue] (dfR) -- (pvhAD);
  \draw[view] (dfC) -- (pvhAG);
  \draw[include] (dfC) -- (dfR);
  \draw[includeleft,thick,blue] (pvhAG) -- (pvhAD);
  \textcolor{blue}{\swpushout[.8]{dfC}{pvhAD}}

  \node[thy] (mvsR) at (6,0) {MvS-Rule};
  \node[thy,thick,blue] (pvhAlt) at (4.2,0) {PvH-Alt};
  \node[thy] (mvsA) at (6,1) {MvS-Aspects};
  \draw[view,thick,blue] (mvsR) -- (pvhAlt);
  \draw[view] (mvsA) -- (pvhAD);
  \draw[include] (mvsA) -- (mvsR);
  \draw[include,thick,blue] (pvhAD) -- (pvhAlt);
  \textcolor{blue}{\swpushout[.7]{mvsA}{pvhAlt}}
  \pic {rightlightning={pvhAM}{pvhAD}};
  \pic {dashedrightlightning={pvhR}{pvhAlt}};
  
  \node at (3,.8) {\footnotesize\ergo \cref{fig:attack}};
  \node at (5,.8) {\footnotesize\ergo \cref{fig:PvHrelevance}};
  \node at (4.2,2.3) {\footnotesize\ergo \cref{fig:PvHdefault}};
\end{tikzpicture}
 \caption{Generating McCarthy's and an alternative Ruling from Legal Theories (see \Cref{fig:PvHview}).}\label{fig:PvHview}
\end{figure}
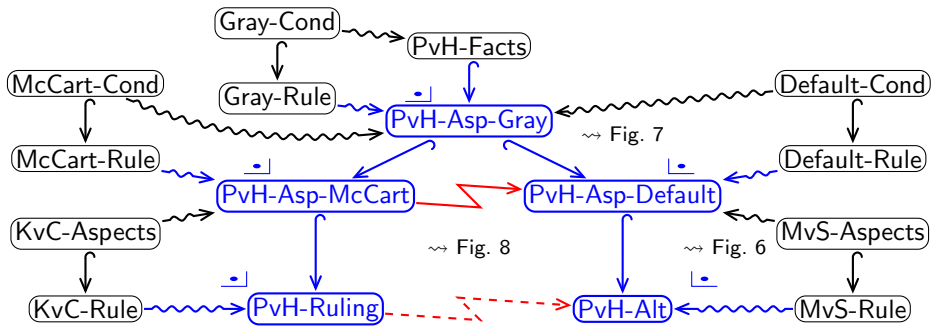

\begin{wrapfigure}r{6.2cm}\vspace*{-1em}
\sf
\begin{tikzpicture}[xscale=.85]
\node (C) at (-2,1.5)   {$\cn{Rule Conditions}$};
\node (C') at (2,1.5) {$\cn{Rule Consequences}$};
\node (E) at (-2,0) {$\cn{Present Case Facts}$};
\node[blue] (E'') at (2,0) {$\cn{Ruling}$};
\draw[view, red](C) -- node[right] {Application} (E);
\draw[view, blue, dashed](C') -- (E'');
\draw[include](C) -- (C');
\draw[include, blue, dashed](E) -- (E'');
\sepushout[.8]{C}{E''};
\end{tikzpicture}
\caption{Application of a rule}\label{fig:application}\vspace*{-1em}
\end{wrapfigure}
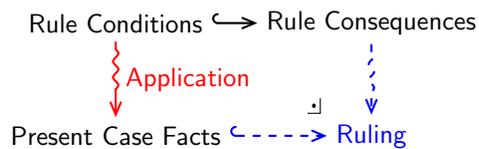

\paragraph{The Running Example continued}
\cref{fig:PvHview} depicts an extension of the formalization of \pvh from \cref{fig:SubsumtionGraph} using analogical reasoning. The black part is that of \cref{fig:SubsumtionGraph}, but the blue part is generated via \textbf{pushouts} (rectangular subgraphs identified by the \textbf{pushout marker} \sepushoutsymb) that correspond to rule/precedent application (see \cref{fig:application} and the discussion there).

\cref{fig:PvHview} contains the backbone of the argumentation of the \emph{Popov v. Hayashi} ruling -- leaving out many details that go beyond introducing the context graph paradigm and thus the scope of this paper.
It starts with the established facts (in the theory \cn{PvH-Facts}) and applies Gray's rule to establish that Popov did not obtain possession of the baseball (and Hayashi did) in \cn{PvH-Aspects-Gray}.
From there, McCarthy would have to apply the ``default rule'':  unless there is a ``proof'' that Popov has possession of the ball we have to assume that he does not.
Subsequently the precedent \mvs (theories \cn{MvS-*})applies to obtain the ruling that Popov has no claim to the ball (\cn{PvH-Alt}; we abstract away from the discussion of ``conversion'').
However he distinguishes the case and instead, introduces a new rule (cf.
\cref{fig:normgraph}) which applied to \cn{PvH-Aspects-Gray} allows him to derive the required proof that Popov has a (qualified) right to possess the ball (theory \cn{PvH-Asp-McCart}).
Finally, this enables him to apply the precedent \emph{Keron v. Cashman} (\cn{KvC-*}) ruling that ownership of the ball is to be split between Popov and Hayashi (theory \cn{PvH-Ruling}).
Note that Judge McCarthy's ruling \cn{PvH-Ruling} contradicts and defeats the default ruling \cn{PvH-Alt} as indicated by the red attack relations.

The three  rectangular patterns marked by ``\ergo Fig. $n$'' in \cref{fig:PvHview} warrant a closer look, we will discuss them in detail below, after we introduce the general principles of views and pushouts in \mmt.  

\paragraph{Views}\label{par:views}
Given theories $S$ and $T$, a \textbf{view} $S\nmmtar{view}\phi T$ states that $T$ is more specific than $S$ after applying the translation $\phi$ from \emph{all} symbols of $S$ to $T$-expressions.
Views have a range of intuitive readings:  
In mathematics, where the notion originates, views are used to formalize examples, interpretation, and even the model relation.
In computer science we can use views to formalize specification refinements and the implementation relations.
In this paper we use it for \textbf{analogy:} $S$ and $T$ are instances of the same (possibly unknown) theory $G$ and $\phi$ an analogy between $S$ and $T$. 

\paragraph{Pushouts}
The category $\mathcal{T}$ of theories and theory morphisms admits colimits and thus \emph{pushouts} -- a special colimit:
Given a theory  $A$ with two morphisms $A \mmtar{include} B$ and $\psi: A\mmtar{view} C$, the \textbf{pushout} of $A$ and $B$ (along $\psi$) is the ``minimal'' theory $P$ that combines $B$ and $C$ identifying material from $A$.

\paragraph{Precedent Application }\label{par:precedent} can be viewed as an instance of analogical reasoning in case law and can thus be formalized as a pushout.
The precedent and the present case are construed as instances of the same abstract rule established by the precedent.
Applying a precedent $A$ to a present case $B$ corresponds to constructing a view $\phi$ from the legal conditions of a rule established in $A$ to the facts of $B$.
In essence, pushouts can be interpreted as a meta-language Modus Ponens rule:
we can apply the ``rule''  $A\mmtar{include}B$ to $C$  if $C$ can be interpreted (i.e. connected with a view) as less specific than $A$.
The pushout $P$ is the result of the ``rule application''. 
\Cref{fig:application} on the right shows the general pattern; it underlies all of the pushouts in~\cref{fig:PvHview} -- with refinements that we discuss below. 

Crucially, the \mmt system can compute pushouts: the information in $A$, $B$, $C$, and $\phi$ fully determines all declarations in $P$~\cite{CMR:colimits:16}; in particular, all names can be generated.

This is the basis for the generated (blue) theories in Figure~\cref{fig:PvHview}. A detailed example is provided by \cref{fig:PvHrelevance}: here the theory \textsf{PvH-Aspects-Default} is subsumed under the rule established by \mvs through the view \textsf{precedent application}. The pushout corresponds to the rule application and creates the ruling: $\vdash \neg \text{Popov has\_claim ball)}$.



\section{Operationalizing Analogical Reasoning in Context Graphs}\label{sec:analogy}
In the pushout model of rule-application in the last section, views operationalize ``analogical transfer'', but the legal literature has identified some additional requirements. In this section we show how these can be modeled.

\paragraph{Analogical Reasoning}
\cite{Stevens2018:RulesvAnalogies} establishes a list of critical questions to analogical reasoning and \cite{AtkinsonBenchCapon2019:RulesvAnalogies} remarks on the difficulty of operationalizing them for a computational model.
\cref{tab:AnalogyRequirements} shows our proposed operationalization of each of these questions.
We will discuss the first three using the \pvh example.
We conjecture that A4 can be handled by considering values (as is done e.g. in \cite{BenchCaponSartor2003:Values}) in addition to rules and prececents but we leave this to future work.

\begin{table}[ht]\footnotesize
  \begin{tabular}{|l|p{5cm}|p{5.8cm}|}\hline
    A1
    &  Are $A$ and $B$ similar in a legally relevant way?
    &  Is there a partial view $\phi$ from  the present case $A$ to a suitable formalization of the precedent $B$?    \\ \hline
    A2
    & Can a successful mapping be made to an aspect of the present-case for every aspect of the precedent-case that the opinion highlights enough to indicate  that it is part of the ratio in the opinion?
    & Does $\phi$'s image in $B$ include all the conditions for the application of a/the rule of $B$      \\\hline
    A3
    & Does the surrounding law allow present-case and precedent-case to be mapped successfully?
    &  Is there any unjustified default in the interpretation of the present case? Does another precedent attack and defeat the currently considered precedent? \\\hline
    A4
    &  Are there no legally relevant differences between A and B?
    &  Is $\phi^{-1}$ onto $A$? If not, is there an $A$-declaration $a$ outside $\phi^{-1}(\phi(B))$ such that $a$ is viewed by an important legal principle.\\\hline
  \end{tabular}
  \caption{Analogical Reasoning according to the Literature}\label{tab:AnalogyRequirements}
\end{table}

\paragraph{A1: Identifying legally relevant similarities}\label{sec:A1}

The first problem in constructing a precedent application consists in identifying which aspects (if any) of the precedent are legally relevant to the present case. This can be done by viewing the precedent from the perspective of the present case  \cite{Stevens2018:RulesvAnalogies}.

We can understand this process in a conservative and constructive way.
Under the conservative interpretation, it corresponds to the fact that views map to \emph{expressions}. In this way, views expose new (derivable) aspects of an \emph{already given} formalization of the precedent that have not been explicitly declared.

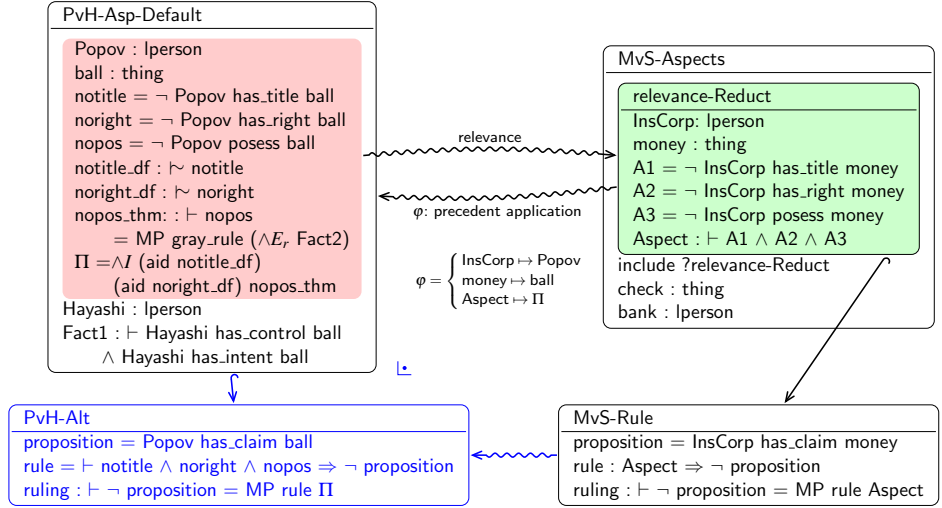
\begin{figure}
  \resizebox{\textwidth}{!}{\sf
\begin{tikzpicture}[remember picture,scale=1.2]
  \node[thy] (pvhAD) at (-.3,4)
  {\begin{tabular}{l}
     PvH-Asp-Default\\\hline\\[-.7em]
     \tikzmarknode[fill=red!20]{pvhADs}
     {\begin{tabular}{l}
        Popov : lperson\\
        ball :  thing \\
        notitle = $\neg$ Popov has\_title ball \\
        noright = $\neg$ Popov has\_right ball \\
        nopos = $\neg$ Popov posess ball\\
        notitle\_df : $\vsim$ notitle \\
        noright\_df : $\vsim$ noright\\
        nopos\_thm: : $\vdash$ nopos\\
        \qquad = MP gray\_rule ($\wedge E_r$ Fact2) \\
        $\Pi$ =$\wedge I$ (aid notitle\_df)\\
        \qquad (aid noright\_df) nopos\_thm \\
      \end{tabular}}\\
     Hayashi : lperson\\
     Fact1 : $\vdash$ Hayashi has\_control ball\\
     \qquad $\wedge$ Hayashi has\_intent ball\\ 
   \end{tabular}};

 \node[thy] (mvsA) at (8,4)
  {\begin{tabular}{l}
     MvS-Aspects\\\hline\\[-.7em]
     \tikzmarknode[thy,fill=green!20]{mvsAs}
     {\begin{tabular}{l}
        relevance-Reduct\\\hline
        InsCorp: lperson\\
        money :  thing \\
        A1 = $\neg$ InsCorp has\_title money \\
        A2 = $\neg$ InsCorp has\_right money \\
        A3 = $\neg$ InsCorp posess money\\
        Aspect : $\vdash$ A1 $\wedge$ A2 $\wedge$ A3\\
      \end{tabular}}\\
     include ?relevance-Reduct\\
     check : thing\\
     bank : lperson 
   \end{tabular}};

 \node[thy,blue] (pvhD) at (.1,0)
  {\begin{tabular}{l}
     PvH-Alt\\\hline
     proposition = Popov has\_claim ball\\
     rule = $\vdash$ notitle $\wedge$ noright $\wedge$ nopos $\Rightarrow$ $\neg$ proposition\\
     ruling : $\vdash$ $\neg$ proposition = MP rule $\Pi$\\
   \end{tabular}};

 \node at (4,2.6) {\footnotesize$\phi=\left\{
     \begin{array}{l}
       \cn{InsCorp} \mapsto \cn{Popov}\\
       \cn{money} \mapsto \cn{ball}\\
       \cn{Aspect} \mapsto \Pi
     \end{array}\right.$};
 
  \node[thy] (mvsR) at (7.7,0)
    {\begin{tabular}{l}
       MvS-Rule\\\hline
       proposition = InsCorp has\_claim money\\
       rule : Aspect $\Rightarrow$ $\neg$ proposition\\
       ruling : $\vdash$ $\neg$ proposition = MP rule Aspect
     \end{tabular}};
  \draw[view] (mvsAs) to[bend left=5] node[below] {\scriptsize $\phi$: precedent  application}(pvhAD);
  \draw[view] (pvhADs) to[bend left=5] node[above] {\scriptsize relevance} (mvsAs); 
  \draw[include,blue] (pvhAD) -- (pvhD);
  \draw[include] ([xshift=-1.5em]mvsAs.south east) -- (mvsR);
  \draw[view,blue] (mvsR) -- (pvhD);
  \textcolor{blue}{\swpushout[.7]{mvsA}{pvhD}}
\end{tikzpicture}}
  \caption{Identifying legally relevant aspects of \mvs.}\label{fig:PvHrelevance}
\end{figure}

Here, a little logical clarification is in order: consider the precedent application view $\phi$ in \cref{fig:PvHrelevance}.
By the theory morphism property, it has to map every constant without definition to an expression in \cn{PvH-Asp-Default}.
The interesting case is the last: in \cn{MvS-Aspects} \cn{Aspect} is an axiom of type $\vdash\cn{A1}\wedge\cn{A2}\wedge\cn{A3}$, so it must be mapped to an (proof)-expression of type $\vdash\cn{notitle}\wedge\cn{noright}\wedge\cn{nopos}$, we call this a \textbf{proof obligation} of $\phi$.
This obligation is fulfilled by the proof term $\phi (\cn{Aspect})=\Pi$. 

Under the constructive interpretation, the precedent's formalization \emph{itself} is to be guided by the present case.
This can be done through a special definitorial variant of a view -- called \textbf{structure} in {\mmt} -- where we import, rename and define content from the present case. This then serves as the starting point for a full formalization of the precedent. Here we will focus on the conservative interpretation.
 
In both cases, formally we have to ensure that there is a (partial) view from the present case's facts to the precedent case's aspects. As we are usually only interested in capturing part of the present case, this view will usually not be total. E.g. deciding whether Popov has a claim to the ball or not as in \cref{fig:PvHrelevance} does not require us to map facts pertaining to Hayashi.

\cref{fig:PvHrelevance} shows how the precedent \mvs which Judge McCarthy used to assess whether Popov can sue for conversion is viewed (conservatively) in light of the facts of \pvh.
In \mvs an insurance company sued a bank over the ``conversion'' of checks issued to non-existing persons by a fraudulent employee and money that had been paid out on receipt of these checks.
The court ruled that the money could not have been converted based on the precedent that depositors have no claim to specific assets in a bank.
In becoming a leading case, the ``rule of the case'' abstracted from this is that ``someone who has no title, possession or right to possession of an object cannot sue for conversion of that object''\cite[p.6]{Atkinson2012:PopovvHayashiRuling}. Judge McCarthy is interested in applying this rule to Popov's situation.
Under the default assumption (cf. \cref{sec:default}) that Popov has no right to possession of the ball, the mapping is onto the application conditions of the rule established by \mvs.
In this way the mapping separates the presently legally relevant aspects from the irrelevant ones (everything concerning the checks, the relationship between the money and the checks, the differences between money and baseballs, etc.).

\paragraph{A2: Totality of Precedent Application }\label{sec:A2}

Next we need to check whether the precedent is \emph{applicable}. This is the case if the reduct of the precedent induced by the present case meets all the application conditions of a rule of the precedent case - as obtains in \cref{fig:PvHrelevance}. There the partial view \textsf{relevance} carves out a reduct of \mvs that suffices to apply the case's rule. This implies that by inverting \textsf{relevance} we get a total view \textsf{precedent application} that allows to map the rule conditions to the present case - thereby enabling rule application via pushout.

However, even partial views from precedents may be useful in the act of distinguishing and with theory construction in ``hard'' cases. The greater the proportion of the precedent that can be mapped, the higher should it be regarded as a candidate for a theory construction starting point. In addition, partial views can serve as heuristics to find complete ones. \mmt is already equipped with a view finder system that can find simple total and partial views. This may help practitioners in finding precedent candidates quickly.

\paragraph{A3: Defaults and Defeasibility}\label{sec:A3}

A precedent application is binding if $\phi$ covers  all ``legally relevant'' aspects of $B$.
Else the court may be justified to \emph{distinguish} $B$ from $A$, that is defeat the application of $A$ to $B$ either by finding a better precedent $A'$ or by setting a precedent itself by adopting a new rule (or, as we see in the next example, both).
Thus analogical precedent application -- just as the application of legal rules -- is defeasible.
On the other hand some rules apply by default if the burden of proof of no other applicable rule can be met. The paradigmatic example is the principle ``innocent unless proven guilty''.
Defaults and defeasibility can be implemented in context graphs through \mmt's \emph{derived modules}. For example, in \cite{KohRapp:cgal19} we implemented \aba in this way. Here we combine this approach with analogical reasoning.


\subparagraph{Default Reasoning}\label{sec:default}

Default reasoning is one of the most important defeasible modes of inference in legal reasoning. The defeasibility of defaults arises from two circumstances: the proof search may not have been exhaustive or new information or rules may have been added to the knowledge base that render formerly unprovable statements provable.

Both situations arise in the law: the first corresponds to a relevant rule from the surrounding law having been ignored; the second arises when new evidence comes to light or a court decides to fashion a new rule. The latter is what happened in \pvh.


Following our approach in \cite{KohRapp:cgal19} we model the default reasoning in \pvh by instantiating \aba. To this end we enrich our metalanguage with an assumption operator $\vsim: bool \longrightarrow type$ and a strict inference rule $aid: \{x:bool\}\vsim x \longrightarrow \vdash x$.

\begin{figure}[ht]
  \sf
\begin{tikzpicture}[yscale=.85]
  \node[thy] (dfC) at (0,2)
  {\begin{tabular}{l}
     Default-Cond\\\hline
     Prop : bool
   \end{tabular}};

 \node[thy] (dfR) at (0,0)
  {\begin{tabular}{l}
     Default-Rule\\\hline
     proposition = Prop\\
     default : $\vsim \neg$ Prop
   \end{tabular}};

 \node[thy] (pvhAG) at (8,2)
  {\begin{tabular}{l}
     PvH-Asp-Gray\\\hline
     \dots\\
   \end{tabular}};

 \node[thy] (pvhA) at (3,1) {PvH-Facts};
 
 \node[thy,blue] (pvhAD) at (8,0)
    {\begin{tabular}{l}
       PvH-Asp-Default\\\hline
       noright\_df : $\vsim \neg$ Popov has\_right ball\\
       \dots\\
     \end{tabular}};

   \draw[include] (pvhA) -- (pvhAG); 
   \draw[view] (dfC) -- node[above] {\scriptsize Prop $\mapsto
     $ Popov has\_right ball} (pvhAG);
   \draw[include] (dfC) -- (dfR);
   \draw[include,blue] (pvhAG) -- (pvhAD);
   \draw[view,blue] (dfR) -- (pvhAD);
   \textcolor{blue}{\sepushout[.60]{dfC}{pvhAD}}
\end{tikzpicture}
  \caption{Inference by default: If Popov's right to possession cannot be proven, it is assumed that he has none.}\label{fig:PvHdefault}
\end{figure}
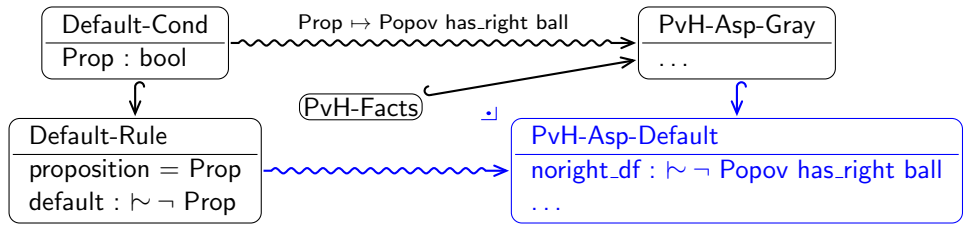

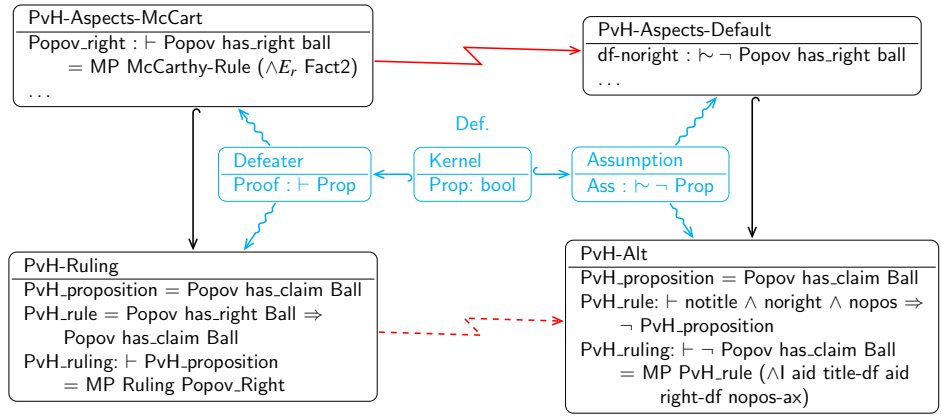
\begin{figure}[h]
  \resizebox{\textwidth}{!}{\sf
\begin{tikzpicture}[remember picture,scale=1.5]
  \node[thy] (pvhAM) at (0,3.2)
  {\begin{tabular}{l}
     PvH-Aspects-McCart\\\hline
     Popov\_right : $\vdash$ Popov has\_right ball \\
     \qquad = MP McCarthy-Rule ($\wedge E_r$ Fact2)\\
     \dots\\
   \end{tabular}};
 
   \node[thy] (pvhAD) at (6.6,3.2)
   {\begin{tabular}{l}
      PvH-Aspects-Default\\\hline
      df-noright : $\vsim\neg$ Popov has\_right ball\\
      \dots\\
    \end{tabular}};
  
  \node[thy] (pvhR) at (0,0) {\begin{tabular}{l} PvH-Ruling\\\hline
                                PvH\_proposition = Popov has\_claim Ball\\
                                PvH\_rule =  Popov has\_right Ball $\Rightarrow$ \\
                                \qquad Popov has\_claim Ball\\
                                PvH\_ruling: $\vdash$ PvH\_proposition\\
                                \qquad = MP Ruling Popov\_Right\\
                              \end{tabular}};
 
                            \node[thy] (pvhD) at (6.6,0) {\begin{tabular}{l} PvH-Alt\\\hline
                                                            PvH\_proposition =  Popov has\_claim Ball\\
                                                            PvH\_rule: $\vdash$ notitle $\wedge$ noright $\wedge$ nopos  $\Rightarrow$\\
                                                            \qquad $\neg$ PvH\_proposition \\
                                                            PvH\_ruling: $\vdash$ $\neg$ Popov has\_claim Ball\\
                                                            \qquad = MP PvH\_rule ($\wedge$I aid title-df aid\\
                                                            \qquad\qquad right-df nopos-ax)\\
                                                          \end{tabular}};

  \draw[include] (pvhAM) -- (pvhR); 
  \draw[include] (pvhAD) -- (pvhD);
  \pic {rightlightning={pvhAM}{pvhAD}};
  \pic {dashedrightlightning={pvhR}{pvhD}};
  \node[color=cyan] (def) at (3.3,2.4){\textsf{Def.}};
  \node[thy,color=cyan] (l1) at (1.2,1.8) {\begin{tabular}{l} Defeater\\\hline Proof : $\vdash$ Prop\end{tabular}};
  \node[thy,color=cyan] (l) at (3.3,1.8) {\begin{tabular}{l} Kernel\\\hline Prop: bool\end{tabular}};
  \node[thy,color=cyan] (l2) at (5.4,1.8) {\begin{tabular}{l} Assumption\\\hline Ass : $\vsim\neg$ Prop\end{tabular}};
  \draw[include,color=cyan] (l) -- (l1);
  \draw[include,color=cyan] (l) -- (l2);
  \draw[view,color=cyan] (l1) -- (pvhAM);
  \draw[view,color=cyan] (l1) -- (pvhR);
  \draw[view,color=cyan] (l2) -- (pvhAD);
  \draw[view,color=cyan] (l2) -- (pvhD);
\end{tikzpicture}}
  \caption{Inheriting an Attack Relation}\label{fig:attack}
\end{figure}

\cref{fig:PvHdefault} shows the situation in which Judge McCarthy finds himself after having inferred via Gray's rule that Popov does not have possession of the ball.
Since he obviously does not have a title (this default is left implicit in the opinion), it remains to check whether he has a right to possession.
In the absence of an applicable rule establishing a right to possession for Popov, it remains to apply the default rule to infer that he has no such right.
Note that the default rule's only condition is that the input variable be a proposition and it creates an assumption indicated by $\vsim$.
Hence anything can be assumed by default and together with the rule $aid$ we may get inconsistent theories.
However, this is unproblematic as such theories are self-attacking (cf. \cref{fig:attack}).

\subparagraph{Defeasibility}



In \pvh Judge McCarthy was unhappy with this default outcome and instead decided to fashion a new rule. McCarthy's rule views \textsf{PvH-Asp-Gray} and via pushout establishes Popov's claim to the ball based on his interrupted efforts to catch it. This new situation now matches the precedent \kvc which he applies. However, his ruling is now in conflict with the default ruling. How can this conflict be resolved?

\cref{fig:attack} shows how McCarthy's rule allows to prove the negation -- technically, the \aba contrary --  of the default assumption \textsf{df-noright}. In theory graphs, attack can be defined \emph{by elaboration} into a subgraph(for details see \cite{KohRapp:cgal19}). In \cref{fig:attack}, the light blue diagram is the elaboration of the red attack arrows: Prop is mapped to \textsf{PvH\_proposition}, \textsf{Proof} to \textsf{Popov\_right} and \textsf{Ass} to \textsf{df-noright}. This corresponds to the \aba definition of attack according to which a set of assumptions $A$ attacks a set of assumptions $B$ if $A$ allows to derive the contrary (here: negation) of an assumption in $B$.

If we apply one of the usual labelings (admissible grounded, preferred, ideal, \dots) to the resulting attack graph, \textsf{PvH\_Aspects\_Default} is defeated and hence the precedent \mvs is distinguished as its conditions cannot be mapped into \textsf{PvH\_Aspects\_McCart}.

\section{Conclusion and Outlook}\label{sec:concl}
We have modeled the well-known case \emph{Popov v. Hayashi} in theory/context graphs, a new, structured, logic-based framework for legal reasoning and argumentation.
We have shown that the mechanism of pushouts is well-suited to  model precedent and rule application and the views involved can adequately capture the  analogical transfer in application.
Moreover, the modular structure of theory graphs nicely captures the intuitive structure of world and legal knowledge
In essence, our approach lifts legal reasoning and argumentation to the theory level and the graph structure can be used to drive the process.
This can be used to foster human understanding of complex legal arguments via the theory graph diagrams and can ultimately support legalTech services like subsumption checking, precedence case finding, and even educational applications in the long run. 


\bibliographystyle{alpha}
\bibliography{kwarcpubs,extpubs,kwarccrossrefs,extcrossrefs,local}
\end{document}